\documentstyle{article}             
\def\beq{\begin{equation}}         \def\eeq{\end{equation}}
\def\bear{\begin{eqnarray}}       \def\eear{\end{eqnarray}}
\def\bt{\begin{tabular}}          \def\et{\end{tabular}}
\def\la{\langle}                  \def\ra{\rangle}
\def\dg{\dagger}                  \def\ci{\cite}
\def\lb{\label}                   \def\ld{\ldots}
\def\lf{\left}                    \def\rt{\right}
                   \def\sm{\small}
                   
\def\nn{\nonumber}

\def\Gam{\Gamma}     \def\k{\mbox{\large$\kappa$}}
\def\Dlt{\Delta}     
\def\sig{\mbox{\large$\sigma$}}

\begin{document}
\baselineskip=15pt

\title{\bf Remarks on the Extended Characteristic Uncertainty 
Relations}
\author{D.A. Trifonov\\
        {\small Institute for Nuclear Research,
        72 Tzarigradsko chauss\'ee, Sofia 1784, Bulgaria}}
\maketitle
\date {}

\begin{minipage}{13cm}
{\small
Three remarks concerning the form and the range of validity of the
state-extended characteristic uncertainty relations (URs) are 
presented. 
A more general definition of the uncertainty matrix for pure and 
mixed
states is suggested. Some new URs are provided. }
\end{minipage}
\vspace{7mm}

In the recent papers \ci{T00,TD} the conventional uncertainty 
relation
(UR) of Robertson \ci{R34} (which includes the Heisenberg and
Schr\"odinger UR \ci{S30} as its particular cases) have been 
extended to
all characteristic coefficients of the uncertainty matrix \ci{TD} and to
the case of several states \ci{T00}. In this letter we present three
remarks on these extended characteristic URs.

The first remark refers to the form of the extended URs \ci{T00}: we 
note
that they  can be written in terms of the principal minors of the 
matrices
involved and write the entangled Schr\"odinger UR \ci{T00} in
a stronger form. The second remark is concerned with the 
extension of the
characteristic inequalities to the case of mixed states and non-
Hermitian
operators. The extension is based on the suitably constructed 
Gram matrix
for $n$ operators and mixed states. The characteristic inequalities 
for
$n$ non-Hermitian operators are in fact URs for their $2n$ Hermitian
components. The last remark is about the domain
problem of the operators involved in the URs. The proper 
generalization of
the uncertainty matrix is suggested as the symmetric part of the
corresponding Gram matrix. For pure states which are in the 
domain of all
product of the operators involved this Gram matrix coincides with 
the
Robertson matrix, and its symmetric part is equal to the 
conventional
uncertainty matrix. \\

Let us first recall the Robertson UR for $n$ observables (Hermitian
operators) $X_1,\ld, X_n$
and a state $|\psi\ra$ and its particular case of $n=2$. The 
Robertson UR
is an inequality for the determinant of the uncertainty matrix $\sig$
(called also dispersion or covariance matrix),
\beq\lb{sig}
\sig_{jk}
= \mbox{$\frac 12$}\la\psi|(X_jX_k+X_kX_j)|\psi\ra -
\la\psi|X_j|\psi\ra\la\psi|X_k|\psi\ra \equiv \sig_{jk}
(\vec{X};\psi),
\eeq
 and it reads
\beq\lb{RUR}
\det\sig(\vec{X};\psi) \,\geq\, \det\mbox{\large$\k$}(\vec{X};\psi)
\eeq
where $\mbox{\large$\k$}(\vec{X};\psi)$ is the  matrix of mean 
commutators,
$\mbox{\large$\k$}_{kj}(\vec{X};\psi) = (-i/2)\la\psi|[X_k,X_j]|\psi\ra$.
For two observables $X$ and $Y$ one has
$\det\sig= (\Dlt X)^2(\Dlt Y)^2-(\Dlt XY)^2$,
$\det\mbox{\large$\k$} = -\la\psi|[X,Y]|\psi|\ra^2/4$,
and the inequality (\ref{RUR}) can be rewritten in the more familiar 
form
of Schr\"odinger UR:
\beq\lb{SUR}
(\Dlt X)^2(\Dlt Y)^2-(\Dlt XY)^2 \,\geq\, \mbox{$\frac 14$}|\la[X,Y]\ra|,
\eeq
where $\Dlt XY$ stands for the covariance of $X$ and $Y$, $\Dlt
XY=\sig_{XY}$, and $\Dlt XX \equiv (\Dlt X)^2$ is the variance of $X$.
Robertson came to his UR by considering the non-negative definite 
Hermitian
matrix $R$,
\beq\lb{R}
R_{jk}= \la\psi\,|\,(X_j-\la X_j\ra) (X_k-\la X_k\ra)\,|\,\psi\ra,
\eeq
which was represented as $R = \sig + i\k$ (the proof of (\ref{RUR}) 
can
be found also in \ci{T99}). In \ci{TD} it was noted that $\det\sig$ and
$\det\k$ are the highest order characteristic coefficients $C_n^{(n)}$
\ci{Gant}  of $\sig$ and $\k$
and Robertson UR (\ref{RUR}) was extended to all the other
coefficients in the form
$C_r^{(n)}\lf(\sig(\vec{X};\psi)\rt) \geq
C_r^{(n)}\lf(\k(\vec{X};\psi)\rt),\quad r=1,\ld,n.
$
In \ci{T00} a scheme for construction of URs for $n$ observables 
and $m$
states was presented. As an example of URs within this scheme 
the
following extended characteristic URs for $n$ observables $X_j$ 
and $m$
states $|\psi_\mu\ra$ were established,

\beq\lb{eur1}
C_r^{(n)}\lf(\mbox{$\sum_\mu$}\sig(\vec{X};\psi_\mu)\rt) \,\geq\,
C_r^{(n)}\lf(\mbox{$\sum_\mu$}\k(\vec{X};\psi_\mu)\rt)
\eeq
\beq\lb{eur2}
C_r^{(n)}\lf(\mbox{$\sum_\mu$} R(\vec{X};\psi_\mu)\rt) \,\geq\,
\sum_\mu C_r^{(n)}\lf(R(\vec{X};\psi_\mu)\rt).
\eeq

Noting that the Robertson matrix in a state $|\psi\ra$ can be 
represented
in the form of a Gram matrix $\Gam^{(R)}$ for $n$ non-normalized
states of the form $||\chi_j\ra = (X_j-\la X_j\ra)|\psi\ra$,\, $\la X_j\ra =
\la\psi|X_j|\psi\ra$,
\beq\lb{Gam^R}
R_{jk} = \la(X_j-\la X_j\ra)\psi\,|\,(X_k-\la X_k\ra)\psi\,\ra
\equiv \Gam_{jk}^{(R)},
\eeq
it was suggested \ci{T00} that Gram matrices for other types of 
suitably
chosen non-normalized states $||\Phi_\mu\ra$ can also be used for
construction of URs of the form (\ref{eur1}) and (\ref{eur2}) for several
observables and states, including the case of one observable and 
several
states. Let us recall that matrix elements $\Gam_{jk}$ of a Gram 
matrix
$\Gam$ for $n$ (generally non-normalized) pure states $||\Phi_j\ra$ 
are
defined as $\Gam_{jk} = \la\Phi_j|\Phi_k\ra$, and $\Gam$ is 
Hermitian and
non-negative definite.  \\

{\bf The first remark}
on the extended URs (\ref{eur1}) and (\ref{eur2}) is that they follow
from a slightly more simple URs in terms of the principal
minors ${\cal M}(i_1,\ld,i_r)$ \ci{Gant} of matrices $\sig$, $\k$ and 
$R$:
\beq\lb{M_r}
\bt{l}
${\cal M}\lf(i_1,\ld,i_r;\sum_\mu\sig_\mu\rt) \,\geq\, {\cal 
M}\lf(i_1,\ld,i_r;
\sum_\mu\k_\mu\rt)$,\\[3mm]
${\cal M}\lf(i_1,\ld,i_r;\sum_\mu R_\mu\rt) \,\geq\, \sum_\mu{\cal
M}(i_1,\ld,i_r;R_\mu)$.
\et
\eeq
The validity  of (\ref{M_r}) can be easily inferred from the proofs of
characteristic URs in \ci{T00,TD}. Let us remind that ${\cal
M}(i_1,\ld,i_n;\sig)=\det\sig$, and the $n$ different ${\cal 
M}(i_1;\sig)$
are equal to the diagonal elements of $\sig$.  URs (\ref{eur1}) and
(\ref{eur2}) can be obtained as sums of (\ref{M_r}) over all minors of
order $r$, since the characteristic coefficient of order $r$ is a sum 
of
all minors ${\cal M}_r$, which here are non-negative. The advantage 
of the
forms (\ref{eur1}) and (\ref{eur2}) is that the characteristic 
coefficients of
a matrix are invariant under the similarity transformations of the 
matrix.
For any Gram matrix and its symmetric and antisymmetric parts 
the
inequalities (\ref{eur1}), (\ref{eur2}) and (\ref{M_r}) are valid.

For two observables $X$ and $Y$ and two states $|\psi_1\ra$ and 
$|\psi_2\ra$
the highest order
inequalities (\ref{M_r}) (the second order) coincide with the 
inequalities
(\ref{eur1}) and (\ref{eur2}) and produce the state-entangled UR (18) 
of
ref. \ci{T00}, which,
after some consideration, can be written in the stronger form
\bear\lb{sheur}
\mbox{$\frac 12$}
\lf[(\Dlt X(\psi_1))^2(\Dlt Y(\psi_2))^2 + (\Dlt X(\psi_2))^2
(\Dlt Y(\psi_1))^2\rt] - \lf|\Dlt XY(\psi_1)\Dlt XY(\psi_2)\rt|\nn \\
\,\geq\,
\mbox{$\frac 
14$}\lf|\la\psi_1|[X,Y]|\psi_1\ra\la\psi_2|[X,Y]|\psi_2\ra\rt|.
\eear
For equal states, $|\psi_1\ra = |\psi_2\ra = |\psi\ra$, the inequality
(\ref{sheur}) recovers the old Schr\"odinger UR (\ref{SUR}). \\

{\bf The second remark}
is, that the extended URs related to any Gram
matrix admit generalizations to mixed states and to non-Hermitian
operators as well.
For $n$ non-Hermitian operators $Z_j$ and a mixed state $\rho$ 
we define a
matrix $\Gam^{(R)}(\vec{Z};\rho)$ as a Gram matrix for the 
transformed
states $\tilde{\rho}_j= (Z_j-\la Z_j\ra)\sqrt{\rho}$ by means of the
matrix elements of the form
\beq\lb{G^R(rho)}
\Gam^{(R)}_{jk}(\vec{Z};\rho) =
{\rm Tr}\lf[(Z_k-\la Z_k\ra)\,\rho\,(Z_j^\dg-\la Z_j\ra^*)\rt].
\eeq
These matrix elements can be represented as Hilbert-Schmidt 
scalar
products $(\cdot,\cdot)_{\rm HS}$ for the transformed states 
$\tilde{\rho}_j$,
\beq\lb{()_HS}
 \Gam^{(R)}_{jk}(\vec{Z};\rho) =
= {\rm Tr}\lf[\tilde{\rho}_k \tilde{\rho}_j^\dg\rt] =
(\tilde{\rho}_k,\tilde{\rho}_j)_{\rm HS},
\eeq
For pure state $\rho = |\psi\ra\la\psi|$,
$(\tilde{\rho}_k,\tilde{\rho}_j)_{\rm HS} = \la (Z_j-\la Z_j\ra)\psi
\,|\, (Z_k-\la Z_k\ra)\psi\ra$.
When a cyclic permutation ${\rm Tr}(Z_k\rho Z_j^\dg) = {\rm 
Tr}(Z)j^\dg
Z_k\rho)$ is possible, then
\beq\lb{G^R(rho)b}
\Gam^{(R)}_{jk}(\vec{Z};\rho) =
{\rm Tr}\lf[(Z_j^\dg-\la Z_j\ra^*)(Z_k-\la Z_k\ra)\,\rho\rt] \equiv
\la (Z_j^\dg-\la Z_j\ra^*)(Z_k-\la Z_k\ra)\ra,
\eeq
and $\Gam^{(R)}(\vec{X};\rho)$ coincides with the Robertson matrix:
$R(\vec{X};\rho)  = \sig(\vec{X};\rho) + i\k(\vec{X};\rho).$

Thus $\Gam^{(R)}(\vec{Z};\rho)$ with elements given by eq. 
(\ref{G^R(rho)}),
is a generalization of the Robertson matrix to the case of non-
Hermitian
operators and mixed states.

For several mixed states $\rho_\mu$ the Gram matrices
$\Gam^{(R)}_{jk}(\vec{Z};\rho_\mu)$, and their symmetric and 
antisymmetric
parts $S(\vec{Z};\rho_\mu)$ and $K(\vec{Z};\rho_\mu)$,
whose matrix elements take the form
\beq\lb{S,K}
\bt{l}
$S_{jk}(\vec{Z};\rho_\mu) = {\rm Re}\lf[{\rm Tr}(Z_k\rho_\mu 
Z_j^\dg)\rt] -
{\rm Re}(\la Z_j\ra^*\la Z_k\ra)$,\\[3mm]
$K_{jk}(\vec{Z};\rho_\mu) = {\rm Im}\lf[{\rm Tr}(Z_k\rho_\mu 
Z_j^\dg)\rt] -
{\rm Im}(\la Z_j\ra^*\la Z_k\ra)$,
\et
\eeq
satisfy the extended characteristic inequalities (\ref{eur1}) and
(\ref{eur2}). We have to note that the characteristic inequalities for
$S(\vec{Z};\rho_\mu)$,\, $K(\vec{Z};\rho_\mu)$ and
$\Gam^{(R)}(\vec{Z};\rho)$ can be regarded as new URs for the $2n$
Hermitian components $X_j$ and $Y_j$ of $Z_j$, $Z_j= X_j+iY_j$.  
The
simplest illustration of this fact is the case of two boson annihilation
operators $a_j= (q_j+ip_j)/\sqrt{2}$ and one state.  For $n=2$ the 
second
order characteristic coefficient $C_2^{(2)}$ is the determinant of the
corresponding matrix. After some consideration we obtain $\det
K(a_1,a_2;\rho) = (\Dlt q_1p_2 - \Dlt q_2p_1)^2/4,$ and
\beq\lb{Sab}
\det S(a_1,a_2;\rho) =
[(\Dlt q_1)^2 + (\Dlt p_1)^2]\,[(\Dlt q_2)^2 +
(\Dlt p_2)^2]/4 - (\Dlt q_1q_2 + \Dlt p_1p_2)^2/4.
\eeq
The characteristic inequality $\det S(a_1,a_2;\rho) \geq \det
K(a_1,a_2;\rho)$ takes the form of a new UR for $q_1,\,q_2,\,p_1$ 
and $p_2$,
\beq\lb{newUR}
[(\Dlt q_1)^2 + (\Dlt p_1)^2]\,[(\Dlt q_2)^2 + (\Dlt p_2)^2]
\,\geq\, (\Dlt q_1q_2 + \Dlt p_1p_2)^2 + (\Dlt q_1p_2 - \Dlt 
q_2p_1)^2.
\eeq
The above consideration suggests that the
symmetric part $S(\vec{X},\rho)$ of $\Gam^{(R)}(\vec{X};\rho)$ 
(defined in
eq. (\ref{G^R(rho)}) with $\vec{Z} \rightarrow \vec{X}$) could be taken 
as
a {\it generalized definition of the uncertainty matrix} for $n$
observables $X_j$ in mixed state $\rho$.
\vspace{3mm}

{\bf The third remark}
concerns the equivalence of the expressions (\ref{G^R(rho)}) and
(\ref{G^R(rho)b})  for the Gram matrix. They are equivalent if the
cyclic permutation of the operators $Z_k\rho Z_j^\dg$ in the trace 
${\rm Tr}
(Z_k\rho Z_j^\dg)$ is possible. For $\rho = |\psi\ra\la\psi|$ this is
rewritten as
$\la Z_j\psi|Z_k\psi\ra = \la \psi|Z_j^\dg Z_k\psi\ra$  and it means 
that
the state $|\psi\ra \in {\cal D}(Z_jZ_k)$, where ${\cal D}(Z_jZ_k)$ is 
the
domain of the operator product $Z_jZ_k$. However it is well known 
that
this not always the case. Example is the squared moment operator 
$p^2 = -
d^2/dx^2$ and any state represented by a square integrable function
$\psi(x)$ which at some points has (first but) no second derivative.
In this sense the expression  (\ref{G^R(rho)}) is more general than
(\ref{G^R(rho)b}). Then the symmetric part $S(\vec{X};\psi)$ of
$\Gam^{(R)}(\vec{X};\psi)$ is to be considered as more general 
definition
of the uncertainty matrix in pure states, and $S(\vec{X};\rho)$, eq.
(\ref{S,K}) with $\vec{Z}=\vec{X}$,\,  -- \, in mixed states. \\

After this work was completed I learned about the very recent
E-print \ci{Chis}, where (in view of the domain problem) the 
expression
${\rm Re}\la X\psi\,|\,Y\psi\ra - \la X\ra \la Y\ra$ is proposed as a
more general definition of the covariance of the Hermitian operators 
$X$
and $Y$ in a pure state $|\psi\ra$.

\end{document}